# A Hybrid Adaptive Educational eLearning Project based on Ontologies Matching and Recommendation System

Vasiliki Demertzi[1], Konstantinos Demertzis[2]

1. International Hellenic University, Department of Computer Science, University Campus, Kavala, Greece, vademer@teiemt.gr;
2. Department of Physics, Faculty of Sciences, International Hellenic University, Kavala Campus, St. Loukas, 65404, Greece; kdemertzis@teiemt.gr;
* Correspondence: kdemertzis@teiemt.gr;

**Abstract:** The provision of the same pedagogical and educational methods to all students is pedagogically ineffective. In contrast, more effectively have proved the pedagogical strategies that adapt to the real individual skills of the students. An important innovation in this direction is the Adaptive Educational Systems (AES) that adjust the teaching content on educational needs and students' skills. Effective utilization of these approaches can be enhanced with Artificial Intelligence (AI) and Semantic Web technologies that can increase data generation, access, flow, integration, and comprehension using the very same open standards that drive the World Wide Web (URIs, HTTP, and HTML). This study proposes a novel Adaptive Educational eLearning System (AEeLS) that has the capacity to gather and analyze data from learning repositories and to adapt these to the educational curriculum according to the student skills and experience. It is an innovative hybrid machine learning system that combines a Semi-Supervised Classification method for ontology matching and a Recommendation Mechanism that uses a sophisticated method from neighborhood-based collaborative and content-based filtering techniques, in order to provide a personalized educational environment for each student.

## 1. Introduction

The World Wide Web (www) today is an unruly construct, with a wide variety of styles. Specifically, last decade, the amount of www content dramatically increased that implies the need to manage and analyze big data volumes, which come from heterogeneous and often non-interoperable sources [1]. The management of these big volumes is further complicated by the need for high-security policies and privacy under the recent General Data Protection Regulation (GDPR) [2]. As the web evolves, the need for semantics technologies that focuses on the importance of the content is an important priority for the research communities. A Semantic Web (SWeb) is basically Structured Data Representation via the combined use of Hyperlinks as Entity Identifier Names, Language for and machine and human-comprehensible sentences/statements using the standard structure and Variety of Notations for creating RDF Language sentences in Documents e.g., RDF-Turtle, JSON-LD, RDF-XML, and others. [3].

Generally, the SWeb technologies "*enable people to create data stores on the web, build ontologies, and write rules for handling data. Linked data are empowered by technologies such as Resource Description Framework (RDF), Sparkle Query Language (SPARQL), Web Ontology Language (OWL), and Simple Knowledge Organization System (SKOS)*" describes to W3C's concept of the web of linked data [4]. Ontologies are an official anthology of terms that used to define an area of interest or to organize the terms that can be used in a domain. They describe potential relations and probable restrictions on employing those terms [5]. With this approach, the search engines will contribute to their more efficient collection and processing of useful web content to the setting up a new global educational system [6].

Modern education promotes teaching and learning through sophisticated methods mainly online. The online learning favors independent learning methods. Online learners must be self-directed towards achieving their academic goals and should be self-motivated. For example, the most popular trend in education for the new era is the Growth Mindset. The idea of a Growth Mindset implies that intelligence evolves through hard work and practice. The implementation of appropriate strategies with open-minded thinking leads to individual and intellectual development in each sector through a process of failures, incentives, and redefinition.

The precipitous evolution of the web and mobile devices has made eLearning adaptable, time-saving, and cost-effective in education process. Besides, since the early days of eLearning, its advantages and have significantly overshadowed those of face-to-face training, making distance education a crucial pillar of every new education and training system [7].

Also, the pandemic of Covid-19 [8] that disrupted the education and training of an entire generation makes necessary the use of eLearning platforms for distance education. The distance education systems use modern communication and information technologies to achieve the essential two-way interaction to accelerate and support the educational process [9]. But the new trends in eLearning philosophy such as interactive videos, learning analytics, mobile-friendly online course platforms, virtual conferences, etc. [10], marks the transition to a new era, that needs to expand the learning process with more sophisticated educational opportunities throughout the life of individuals. The ternary relationship that develops between the instructor, the trainee, and the educational material replaces the dual relationship between the instructor and the trainee that until now characterized conventional education [11].

Simultaneously, the rapid development of the cloud computing, the SWeb methodologies, and especially the AI technologies, offer new opportunities in the future development of innovative systems that will allow the smarter management of learning content, for providing personalized educational environments [12].

The SWeb technologies are as much about the data as they are about rational and logic but does not agreement with amorphous content. It is about representative not only organized data and links but also the implication of the main theories and relations. For example, the RDF is the introductory technology in the SWeb stack, which is an adaptable graph information prototype that does not entail rationality or interpretation in any way. Even the elements of the SWeb stack that arrangement with interpretation and assumption are prepared in well-understood official semantics and can usually be conveyed via straightforward sets of instructions [5]. As such, they lack both the complication and the vagueness of AI methods that are based on machine learning and neural prototypes.

AI defined as "a system's ability to correctly interpret external data, to learn from such data, and to use those learnings to achieve specific goals and tasks through flexible adaptation" [13]. Also, an AI system includes capabilities to learn from experience and connectivity and can adapt according to the current situation.

The most important developments concerning the combination of AI and SWeb in education and more specifically in the modern eLearning systems focus on:

1. in information management with appropriate ontologies for optimized performance. The use of ontologies in collaborative environments where collective content are produced, will allow correlations between heterogeneous sources (documents, emails, etc.) in order to easily retrieve all the absolutely relevant information.

2. in the digital libraries where they need to comply with the semantic ontologies and organize their librarian catalogs in a semantic way so that search engines can locate the appropriate content.

3. In the development of innovative applications and eLearning platforms, which using semantic ontologies, will allow the transform of distance education, creating friendly in search engines semantic "maps" of learning material and content.

AES, accepting the above wording, are new technologically supported education systems that adapt the provided educational content to the specific educational needs of each trainee or group of trainees in order to achieve sophisticated learning [6]. They also provide specialized support to the trainees taking into account the learning needs, the special characteristics of learners in addition to their evolution during their study [10].

The contribution of the SWeb and ontologies matching technologies, and especially the artificial intelligence in the development of a novel eLearning architecture, is the motivation of this paper. Specifically, this paper proposes a novel AEeLS, which with extensive use of AI methods, allows the modeling of the process of retrieval and management of information based on semantic criteria, for the needs of individualized education of each student.

The sections appear in the rest of the paper in the following prescribed order as follows: Section 2 presents the related work about the applicable AES that have used AI models. Section 3 illustrates the suggested prototype and describes the methodology, section 4 presents the dataset used and the outcomes of the proposed algorithmic approach and definitively, section 5 contains the conclusions.

## 2. Related Work

Online collaborative has highlighted the eLearning approaches as an essential part of modern educational system. Universities, organizations, and companies have adopted eLearning as a more flexible and effective way to train their students, executives, or employees. However, the current and future trends in eLearning prove that it is a field for continuous innovation and research.

In this paper [14], the authors have presented a complete review of ontology-based recommendation for e-learning. The impact of the work is two-fold. First, they have abridged the research accomplishments in the area of ontology-based recommenders from 2005 to 2014 by organizing the manuscripts according to the year of publication and classifying with emphasis on recommendation methods, knowledge representation, ontology categories, ontology representation language, and recommended learning possessions. Secondly, they have given a complete review of the future tendencies on the ontology-based recommendation for e-learning.

There are some scientific papers, associated to numerous issues applicable to the advancement AEeLS of the present work. For example, the research [15] discovers several tactics for learning metadata mining, whose one of the most valuable open challenges is the recognition of Learning Objects and the metadata that can be gained from them. Also, both Mao et al. [16] and Liu et al. [17] demonstrate how Ontology Matching can be specified as a binary classification problem, forcing use of most well know machine learning algorithms. In the earlier work, an approach for locating relations among two ontologies using Support Vector Machines (SVM) is introduced. The investigational findings show promising are remarkable when contrasted compared to additional mapping techniques.

In addition, the paper [18] propose a novel ontology matching method that uses again SVMs, demonstrating a precision of the order of 95% in their investigational outcomes. Also, in the [19] research, the authors have suggested an ontologies method for the educational domain modeling. They have explained in detail how to build e-Learning ontologies and how they are demoralized in order to express and implement personalized e-Learning practices based on Grid Technologies and several educational methods are applied in order to improve the e-Learning experiences.

Other research work [20], explore the ontology mapping problem based on concept classification by decision trees algorithms that introduces a similarity measure among two portions fitting to distinct ontologies. Nonetheless, the effort does not give analytical precision results, although claiming that the method produced is speedier at implementation due to the less evaluations required.

A different approach presented by the [21] that introduce a graph-based semantic explanation method for improving instructive content with linked records, to gain information exploration with superior recall and precision.

Metaheuristics have also had an important role in the vicinity of e-learning. In this sense, Luna et al. [22] propose a novel concept for finding studying rules applying evolutionary metaheuristic procedures.

Moreover, Peñalver-Martinez et al. [23] employ some natural language processing methods to content produced for attitude mining with remarkable results.

Also, Wang et al. [24] presents a classification method for less widespread webpages based on suppressed semantic analysis and difficult set patterns for the automated tagging of web pages with related content.

An automatic document classifier system based on ontology and the naive Bayes classifier is proposed in the paper [25]. The main concept is to first establish a keyword synonymous table by experts for narrowing down the range and getting the consistency of keywords. The formal concept analysis is then used for establishing knowledge ontology through the complex categories and attributes relation. Finally, the ontology is applied to a naive Bayes classifier to get the automatic document classifier system.

Also, there are several research in the area of ontologies that using data mining techniques [26] and machine learning algorithms such as neural networks [27][28], K-nearest neighbor (KNN) [29][30] and Support Vector Machine (SVM) classifiers [31]. In addition, aiming at the recommendation accuracy of user-item rating matrix sparse generation many research proposed the use of Collaborative Filtering (CF) recommender algorithm [32][33][34].

On the other hand, the investigation of smart recommendation systems, have noticed great recognition and usage in e-market systems. Though, authors of [35] introduce an online curricula recommendation system, which joins numerous clustering methods in order to prove that machine learning approaches can enhance significant the estimation procedure of lessons engaged in e-learning ecosystems.

Also, Gladun et al. [36], introduces a multi-agent recommendation system for automated response relating to expertise achieved by learners in e-learning programs, holding improvement of the SWeb technologies.

Finally, other research methods on distance learning are concentrated on recommending a narrative approach of microlecture via mobile technologies and web platforms, whereas others centered on developing educational perspectives [37].

## 3. Methodology

Because eLearning structures' methodology is an exceedingly complicated method, trainers cannot be centered only on the use of pathetic insulated content and inventions based solely on the old and maybe obsolete educational materials. The content classification based on the student needs, should not be a labor-intensive and time-consuming procedure, something that will introduce an critical disadvantage to the education system. Perspective, the use of additional efficient techniques of education supervision, with abilities of automatic monitor the educational content and use of specific materials for every student is important to every modern educational system.

It is also important the update the eLearning philosophy and its transformation into an Adaptive Educational eLearning System. The ideal AEeLS includes advanced AI methods for real-time scrutiny of the educational needs both known and unknown students, instantaneous reports, statistics visualization of progress, and other sophisticated techniques that maximize the education experience alongside with fully automated content evaluation process by semantic technologies.

Dissimilar to other methods that have been suggested in the literature concentrating on static tactics [18][20], the dynamic prototype of AEeLS produce an evolving educational tool without special needs and hardware resources requirements.

The algorithmic approach of the suggested AEeLS comprises in the first stage an Ontologies Matching process from www in order to find the relevant educational content as you can see in the illustration of the proposed model, in Figure 1. In the second stage, the content checked for the precision and accuracy and a Recommendation Mechanism proposes new relevant material in order to produce an extremely fitted curriculum for each student (stage 2 in Figure 1). The following Figure 1 is a depiction of the suggested AEeLS prototype:

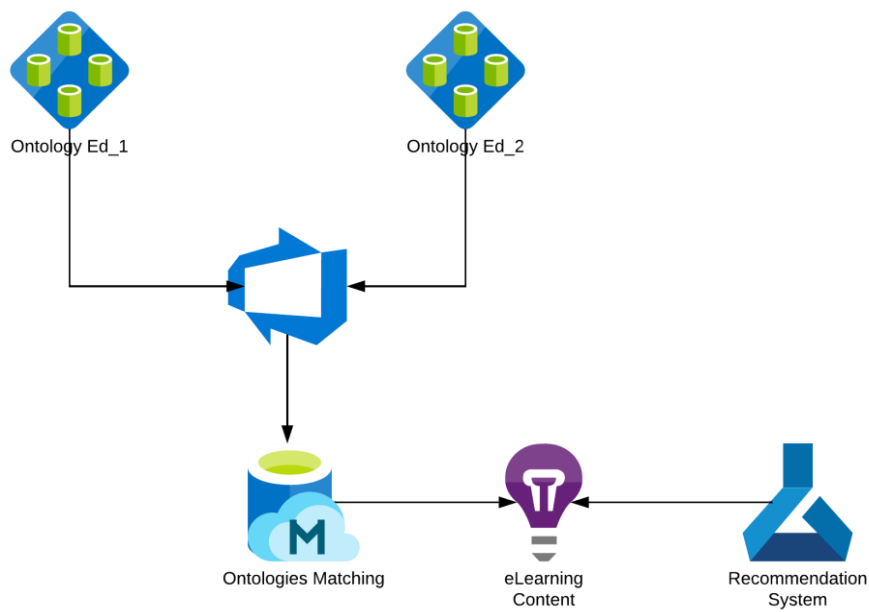

Figure 1. AEeLS model.

*3.1 Ontologies Matching*

The ontologies are a formal structured information framework and a clear definition of a common and agreed conceptual formatting of possessions and interrelationships of the objects that actually exist in a specific area of interest. The main components of the ontologies are classes, properties, instances and axioms. Classes exemplify adjusts of objects within a specific area. Properties define the various characteristics of theories and constrictions on these characteristics. Both of them can be formed into separate hierarchies. Instances represent the concepts and axioms are proclamations in the form of logic to constrain values for classes or properties [38].

Officially an ontology can be defined as below [39]:

$$O=\{C, P, H^C, H^P, I, A^O\} \quad (1)$$

where $C$ and $P$ represent classes and properties, $H^C$ and $H^P$ are the hierarchy of them, $I$ is a set of instances and $A^O$ is a set of axioms.

The proposed Ontologies Matching Mechanism (OMM) based on advanced computational intelligence and machine learning techniques. The purpose is to develop a fully automatic technique for extracting information and controlling the effectiveness of student needs [40]. In particular, this subsystem automates the extraction, analysis, and interconnection of educational web content material based on relevant ontologies for further processing. It also allows for the effective detection of contradictory instructions or content interrelated to the transmission of the particular information to certify that they cannot be used to the disorientation of learning purposes. To achieve this, ontology matching techniques using AI methods used.

Ontology matching is a hopeful method of the semantic heterogeneity dilemma. It uncovers correlations among crucially linked knowledge entities of the ontologies. These correlations can be applied for innumerable tasks, such as ontology integration, query responding, and data conversion. Thus, matching ontologies allows to interoperate and also to information transfer and data integration in the paired ontologies [41].

The aim of ontology matching is the procedure of establishing correlations among conceptions in ontologies to arise an arrangement between ontologies, where an arrangement contains a set of correlations amongst their rudiments so that significant similarity can be equivalent. Given two ontologies $O_S$ (source ontology) and $O_T$ (target ontology) and an entity $e_s$ in $O_S$, the procedure ontology matching $M$ denoted as a process that find the entity $e_t$ in $O_T$, that $e_s$ and $e_t$ are deemed to be equivalent [42].

It should be emphasized that the ontology matching process it can be subsumption, equivalence, disjointness, part-of or any user specified relationship. The most significant matchings or alignments can be categorized in three particular sections [43]:

1. Similarity vs Logic: This category concerns the similarity and logical equivalence among the ontology terms.
2. Atomic vs Complex: With regard to that category the alignment considers if it is "one-to-one", or "one-to-many".
3. Homogeneous vs Heterogeneous: In the third category, the alignments examines if it is on terms of the same type or not (e.g., classes to classes, individuals to individuals, etc.).

Usually, an ontology matching tactic applies numerous and different categories of matchers such as labels, instances, and taxonomy forms to recognize and estimate the resemblance between ontologies. The easiest strategy is to aggregate the similarity standards of each object pair in a linear prejudiced mode and decide on a suitable threshold to recognize matching and non-matching pairs. Though, given a matching condition, it is difficult to define the right weights for each matcher [44]. In recent past, many ontology matching approaches and weighting strategies have been suggested to adaptively verify the weights such as Harmony [44] and Local Confidence [45], but there is no single strategy.

Against, the machine learning based ontology matching methods have been proved to get more precise and reliable matching consequences [46]. Specifically, the supervised machine learning methods use a set of validated matching pairs as training instances, in order to apply a learning patterns strategy that can be find the accurate matches from all the applicant matching pairs. On the other hand, the unsupervised machine learning methods uses arbitrary and heuristic strategies to matching pairs without orderly and modeled methodology. Comparing the machine learning approaches, supervised methods usually get better results [46].

However, the main weakness of the techniques with full supervision is that they need a substantial amount of labeled training examples to create a prognostic system with acceptable performance. The training dataset is mostly accomplished by hand instructor, which is a difficult and inefficient procedure. In addition, the current method only give the comparison values purely as numeric features, without taking their critical appearances into account [47].

As an alternative, the key characteristic of training with Semi-Supervised technique is the creation of the robust prototype with the usage of pre-classified sideways with unlabeled instances. This tactic works on the situation that the input patterns with and without labels, belong to the similar marginal distribution, or they follow a mutual formation. Largely, unlabeled data offer valuable evidence for the discovery of the whole dataset data structure, though separately the arranged data are presenting in the learning procedure. Thus, even the most thoughtful real-world complications can be developed successfully, based on the crucial oddities that describe them [47].

The OMM uses a semi-supervised learning ontology matching innovative method in order to take advantage of a small set of labeled entity pairs to enhance the training procedure. The technique first utilizes the central relationships in the resemblance area and after receiving more training instances, it classifies the rest entities pairs into matched and non-matched classes. Finally, the suggested method define a new set of constrictions to adapt the probability matrix in the labeling process, which help to increase the performance of matching outcomes [48].

The semi-supervised learning method is suitable for the OMM as ensures high-speed, vigorous and efficient classification performance. Moreover, it is easily adjustable and applicable method. Also, it is a pragmatic machine learning technique that can model the ontologies matching challenge based on a section of few pre-classified data vectors, exposing the relationships amongst the taxonomy constructions of ontologies [47-48].

Specifically, the OMM applies a hybrid algorithmic approach that combines the naive Bayes classifier, Collective classification that is a combinatorial optimization method, and fuzzy c-means clustering algorithm in order to produce a quicker and more elastic combined Fuzzy Semi-Supervised Learning scheme. The most significant novelty and improvement of the suggested method is the easy validation of the classification procedure for a first time seen data, based on vigorous calculable features. The theoretic contextual of the system's core is offered in the next subsections.

The naive Bayes classifier [49] is an applied learning technique based on a probabilistic demonstration of a data structure, representative a set of random variables and their suppositious individuality, in which complete and shared probability distributions are validated. The impartial of the procedure is to classify an example X in one of the given classes $C_1, C_2, .., C_n$ by a probability model well-defined rendering to the model of Bayes theorem. These classifiers make probability valuation rather than predicting, which is frequently more beneficial and operative. Here the forecasts have a score and the determination is the minimization of the probable rate. Each class is characterized by a prior probability.

We make the supposition that respectively example X belongs to a class $C_i$ and based on the Bayes theory we estimate the posteriori probability. The measure P relating a naive Bayes classifier for a set of examples, expresses the probability that $c$ is the value of the dependent variable $C$, based on the values $x=(x_1, x_2, ..., x_n)$ of the properties $X=(X_1, X_2,..., X_n)$ and it is given by the subsequent equation 2 where the feature $x_i$ is measured as independent [49]:

$$P(c|x) = P(c) \cdot \prod_{i}^{n} P(x_i|c) \quad (2)$$

The estimation of the above amount for a set $N$ instances is done by using the equations 3, 4 and 5:

$$P(c) = \frac{N(c)}{N} \quad (3)$$
$$P(x_i|c) = \frac{N(x_i,c)}{N(c)} \quad (4)$$

For a typical $x_i$ with distinct values, the Probability is projected by equation 5.

$$P(x_i|c) = g(x_i, \mu c, \sigma c2) \quad (5)$$

where $N(c)$ is the number of instances that have the value c for the depended variable, $N(x_i,c)$ is the number of cases that have the values $x_i$ and c for the characteristic $X_i$ and the depended parameter individually and $g(x_i, \mu c, \sigma c2)$ is the Gaussian probability density function with an average value $\mu c$ and variance $\sigma c$ for the characteristic $x_i$.

Collective classification [50] is a combinatorial optimization method, in which we are providing a set of connections, $V = \{V_1, ..., V_n\}$ and a neighborhood function $N$, where $Ni \subseteq V \setminus \{Vi\}$. Each node in $V$ is an undiscriminating variable that can take a value from an appropriate area. $V$ is supplementary separated into two sets of nodes: $X$, the experiential variables and $Y$, the nodes whose values need to be defined. Our task is to label the nodes $Y_i \in Y$ with one of a small amount of labels, $L = \{L_1, ..., L_q\}$; we'll use the shorthand $y_i$ to infer the label of node $Y_i$.

Similarly, according to Zadeh [51] each element "$x$" of the universe of dissertation "$X$" fits to a Fuzzy Set (FS) with a degree of membership in the closed interval [0,1]. Thus, the following function 6 is the mathematical base of a FS [51]:

$$S = \{(x, \mu s(x)/\mu s: X\{[0,1]: x\} \mu s(x)\} \quad (6)$$

The next equation 7 is an occasion of a normal Triangular Fuzzy Membership Faction (FMF). It must be clarified that the "$a$" and "$b$" parameters have the values of the lower and upper bounds of the raw data independently [51]:

$$\mu_s(X) = \begin{cases} 0 \text{ if } X < a \\ (X-a)/(c-a) \text{ if } X \in [a,c) \\ (b-X)/(b-c) \text{ if } X \in [c,b) \\ 0 \text{ if } X > b \end{cases} \quad (7)$$

Rendering to the typical (crisp) classification methods, each example can be allocated only to one class. Thus, the class membership value is either 1 or 0. In general, classification approaches decrease the dimensionality of a multifaceted datasets by grouping the data into a set of classes. On the other hand, in fuzzy classification, an example point can be allocated to numerous classes with a dissimilar degree of membership. The fuzzy c-means clustering procedure primarily gives random values to the cluster centers and then it assigns all of the data vectors to all of the clusters with varying Degrees of Membership (DoM) by calculating the Euclidean distance.

The Euclidean distance of each data point $x_i$ from the center of each cluster $c_1 \ldots c_j$ is intended based on equation 8 [52].

$$d_{ji} = \|x_i - c_j\|^2 \qquad (8)$$

where $d_{ji}$ is the distance of $x_i$ from the center of the cluster $c_j$. Then the DOM of each data point to each cluster is estimated based on equation 9:

$$\mu_j(x_i) = \frac{\left(\frac{1}{d_{ji}}\right)^{\frac{1}{m-1}}}{\sum_{k=1}^{p}\left(\frac{1}{d_{ki}}\right)^{\frac{1}{m-1}}} \qquad (9)$$

where $m$ is the fuzzification constraint with values in the interval [1.25,2] [40]. The values of $m$ stipulate the degree of overlapping among the clusters. The defaulting value of $m$ is equal to 1.2. The process has the succeeding direct constraint in the DOM of each point [29]. See equation 10 [52]:

$$\sum_{j=1}^{p} \mu_j(x_i) = 1 \quad i = 1,2,3,\ldots k \qquad (10)$$

where $p$ is the amount of the clusters, $k$ is the amount of the data points, $x_i$ is the $i$-th point and $\mu_j(x_i)$ is a function that proceeds the degree of membership of point $x_i$ in the $j$-th cluster $i=1,2,\ldots.k$. Then the centers are estimated again.

The subsequent equation 10 is used for the re-calculate of the values of new cluster centers [52]:

$$c_j = \frac{\sum_i [\mu_j(x_i)]^m x_i}{\sum_i [\mu_j(x_i)]^m} \qquad (11)$$

where $c_j$ is the center of the $j$-th cluster with ($j=1,2\ldots.p$), and $x_i$ is the $i$-th point [52]. This is an iterative system and the whole procedure is repeated till the centers are stabilized.

The OMM is an advanced hybrid method based on the amalgamation of soft computing tactics. Let us deliberate a supervised learning situation with a training set of size N $\{X,Y\}$ = $\{x_i, y_i\}_{i=1}^{N}$, where $x_i \in R^{n_i}$ and $y_i$ is a binary vector of size $n_o$. It must be clarified that $i$ and $n_o$ are the dimensions of the input and output respectively.

The OMM primarily achieves Semi-Supervised Clustering (SSC). This earnings that cluster assignments may be already known for some subset of the data. The final aim is the classification of the unlabeled observations to the appropriate clusters, using the known assignments for this subset of the data. At the same time the procedure produces the degree of membership of respectively record to its cluster.

The clustering validation procedure is accomplished by engaging the "*classes to clusters*" (CL_A_U) technique, that accepts SSC. Formerly a minimum data sample is used covering of the clusters resulting from the SSC development (labeled data). The residual unlabeled data are used to dynamically arrangement and regulate the classes based on their DOM.

Essentially, the CL_A_U method consigns classes to the clusters, based on the popular value of the class quality within each cluster. The class quality is preserved like any other feature and it is a part of the input to the clustering procedure.

The objective is the valuation as to whether the designated clusters match the quantified class data. In the CL_A_U evaluation, you tell the scheme which characteristic is a prearranged "class."

Then this is detached from the data before transient to the SSC procedure. The CL_A_U evaluation, finds the minimum error of mapping classes to clusters (where only the class labels that match to the examples in a cluster are measured) with the restriction that a class can only be mapped to one cluster.

The arisen classes are fuzzified by conveying them appropriate Linguistics, in order to get a accurate consistency among the related standards of the dataset under study.

The whole procedure is obtainable in the Algorithm1 underneath.

| Algorithm 1. The OMM Algorithm |
|---|
| **Inputs**: Input labeled data $D_l$, clusters of the labeled data $L_l$ and a set of unlabeled data $D_u$ |
|     **Stage 1**: *% Initialization of clusters* |
|         Recognize the separate number of clusters based on $L_l$ |
|         For each cluster, produce matrices with the mean and standard deviation of all $D_l$ |
|     **Stage 2**: *% Estimate the new centers of the clusters* |
|         For every cluster, reconstruct these matrices, based on the testing data $D_u$ |

    Estimate a variable, based on the formula below:
$$x = (1./(2*pi*ns.^2)).*exp(-((test-nm).^2)./(2.*sn.^2))$$
   where *ns* is the new standard deviation matrix, *nm* is the new mean matrix and test $D_u$
    Sum all these variables for each cluster
  **Stage 3**: *% Estimate the winner cluster for each record*
    For every testing data $D_u$, find the minimum value of the summary calculated beforehand.
     *% Estimate the fuzzy membership values for every cluster for every record*
    For every testing data $D_u$ and for every class, divide the mean matrix with the sum of the values intended before (normalization probability – membership value)
**Outputs**: Winner cluster for each testing data $D_u$, $C_u$ and fuzzy membership values for every cluster
    for every testing data $D_u$, $F\_M\_V_{u,j}$ (*j* the number of clusters)
  **Stage 5**: *% Validation of the clustering process*
    Repeat Stages 1 – 3 from the previous portion, only this time from $D_u \rightarrow D_l$, using $C_u$ as labels
**Output**: Winner cluster for each testing data $D_l$, $L2_l$
  **Stage 6**:
    For every primarily labeled data $D_l$:
    Compare the preliminary label $L_l$ with $L2_l$
    Create confusion matrix based on these comparisons
  **Stage 7**:
    Repeat Stages 5 - 6 for every $D_w$ of $D_u$
  *% Generalization of the amount of the extreme suitcases, based on the fuzzy membership values*
**Inputs:** The winner class for every record ($C_u$) and the fuzzy membership values for each record
    ($F\_M\_V_{u,j}$)
  **Stage 8**:
    For every record:
    *If $max(F\_M\_V_{u,j}) = A$ AND $F\_M\_V_{u,A} - max2(F\_M\_V_{u,j}) \le$ threshold,* then
     *% $max2(F\_M\_V_{u,k}) = k$, the second biggest membership value*
    Modification the winner class for this record to *k* ($C_u = k$)
**Outputs**: Updated winner cluster for each record $C_u$

In conclusion, the proposed algorithm initially performs clustering using a small number of labeled data, in order to categorize a number of records in clusters. Typically, every cluster is assigned a characteristic center of gravity value. During iterations, the values of the centers are adjusted and when the center stabilizes, the iterations are terminated. Initially, a minimum data sample related to the obtained clusters (labeled data) is used. The remaining unlabeled data which ignore the class attribute are used to provide useful information related to the structure of the overall data set, as they dynamically modulate and adjust the classes based on the values that belong to each cluster.

*3.2. Recommendation Mechanism*

  The Recommendation Mechanism (RMm), is a machine learning method [53] in the AEeLS to create intelligent rules for intervention decisions and offer personalized real-time information for the students educational needs with Collaborative Filtering (CF) [54] technique.

  CF is a machine learning method of making filtering about the conception by accumulating preferences or unique information from several users (collaborating). In the more general sense, CF is the method of filtering for data or outlines using procedures affecting collaboration between various agents, opinions, information resources, etc. Usually, a workflow of a CF can be defined as below [54]:
1. A user extracts the predilections by ranking objects of the structure. These grades can be considered as an estimated description of the user's importance in the related area.
2. The scheme match up this user's rankings compared to other users' and discovers the individuals with most "related" preferences.
3. With similar individuals, the method indorses substances that the comparable operators have ranked highly but not yet being ranked by this individual.

CF systems are separated in memory-based and model-based methods [54]. The most useful technique for this purpose is to allocate weight to the impacts of the neighbors, so that the nearer neighbors provide more to the average than the more distant ones [55]. In addition, CF methods include cluster-based approaches [56], Bayesian techniques [57], Pearson correlation processes, vector similarity practices, regression strategies and error-based tactics [58]. Currently, CF methods have been applied to many kinds of systems including recognizing and observing applications, environmental sensing over large areas, financial process and electronic commerce and web applications [55][58].

Traditional CF methods face two major challenges: data sparsity and scalability [55]. In the RMm, we use a hybrid technique from neighborhood-based CF and content-based filtering that addressing these challenges and improve quality of recommendations [56].

The aim of this hybrid method trying to attain more tailored intellectual directions for intervention decisions and personalized recommendation in real-time information for the student's educational needs based on skills. This hybrid technique is more adaptable, in the sense that they can be applied to heterogeneous ontologies and with some care could also provide cross-domain recommendations. Also, it works greatest when the operator space is enormous, it is easy to implement, and it scales well with no-correlated substances and does not need multifarious modification of parameters [59].

*3.3. Performance Metrics*

In this research the classification performance is valued by the usual evaluation procedures: Precision (PRE), Recall (REC) and F-Score indices that are well-defined as in calculations 12, 13 and 14 correspondingly [60-61]:

$$\text{PRE} = \frac{\text{TP}}{\text{TP} + \text{FP}} \quad (12)$$

$$\text{REC} = \frac{\text{TP}}{\text{TP} + \text{FN}} \quad (13)$$

$$\text{F} - \text{Score} = 2X \frac{\text{PRE X REC}}{\text{PRE} + \text{REC}} \quad (14)$$

Also, the validation method used the 10-fold cross-validation method because the quantity of available examples is relatively larger, which in turn bargains statistically sound performance capacities [60-61].

The testing hardware and software conditions for all simulations are listed as follows: PC Intel Core I7-10700K 3.80GHZ CPU, 64GB DDR4-2933 RAM, Ubuntu 18.04 LTS, Anaconda TensorFlow (Python).

**4. Dataset and Results**

The suggested hybrid model was certified through examinations, which were done on datasets engaged from the Ontology Alignment Evaluation Initiative (OAEI) 2014 [62] operation, as well as on data occupied from two well-known educative content repositories: ADRIADNE [63] and MERLOT [64]. Thus, two datasets were constructed, covering patterns representative the relations among pairs of Learning Objects engaged from two dissimilar ontologies absorbed in the Open and Distance Learning context.

For the first experimental test rendering the [63], the OAEI 2014 dataset was used, for responsibility the problem of Instance Matching Track, more accurately for the Identity Recognition Task [62] and specifically is to find an appropriate similarity function, in order to build pairs of objects which are actually close in significance. Through the passable use of a given resemblance purpose, the ontologies matching problem transformed into a binary pattern classification problem.

The next trial contains on doing a match among two diverse educative content repositories (ADRIADNE and MERLOT) in Learning Objects Metadata arrangement, based on a sample of 100 from each repository, associated to the Computer Sciences subject.

The ADRIADNE Foundation obtainable a provision that is the ability to convert the metadata of the substances into well-known stipulations, such as Learning Objects Metadata and Doublin Core.

MERLOT is one of the principal open access warehouses for educative topics and is shaped for use by research communities. Comprises a congregation of learning assets and educational resources, such as: animations, case studies, collections, questionnaires, simulators, etc.

In this experimentation according the [63], a total of 100 1:1 matching instance were created from both ontologies. The features extraction takes into account for the pattern structure: title, description, keywords, and type of resource [65].

The following table 1, presents a wide evaluation for both datasets, by engaging competitive methods namely: Radial Basis Function Neural Network (RBFNN), Group Method of Data Handling (GMDH), Polynomial Neural Networks (PNN), Feedforward Neural Networks using Genetic Algorithms (FFNN-GA), Feedforward Neural Networks using Particle Swarm Optimization (FFNN-PSO), SVM and Random Forest (RF).

Table 1. Comparison between algorithms (1st experimental test)

| OAEI 2014 data bank | | | |
|---|---|---|---|
| Classifier | PRE | REC | F-Score |
| **OMM** | **0.904** | **0.908** | **0.906** |
| RBFNN | 0.710 | 0.700 | 0.709 |
| GMDH | 0.845 | 0.846 | 0.848 |
| PANN | 0.813 | 0.818 | 0.817 |
| FFNN-GA | 0.887 | 0.888 | 0.889 |
| FFNN-PSO | 0.891 | 0.889 | 0.892 |
| SVM | 0.895 | 0.897 | 0.897 |
| RF | 0.900 | 0.900 | 0.901 |

Table 2. Comparison between algorithms (2nd experimental test)

| ADRIADNE and MERLOT | | | |
|---|---|---|---|
| Classifier | PRE | REC | F-Score |
| **OMM** | **0.981** | **0.981** | **0.982** |
| RBFNN | 0.888 | 0.889 | 0.889 |
| GMDH | 0.940 | 0.942 | 0.946 |
| PANN | 0.901 | 0.902 | 0.902 |
| FFNN-GA | 0.963 | 0.962 | 0.962 |
| FFNN-PSO | 0.965 | 0.964 | 0.964 |
| SVM | 0.976 | 0.977 | 0.976 |
| RF | 0.975 | 0.976 | 0.978 |

Tables 1 and 2 demonstrates obviously that the suggested technique has greater performance for both datasets which is relatively promising contemplating the complexities faced in this problem. It is crucial to say that evaluating several factors that can define a type of challenge discussed here is a partially individual non-linear and dynamic process.

# 5. Conclusions

*5.1 Discussion*

In this paper proposed a hybrid [66-69], sophisticated [70], dependable [71-72] and vastly effective eLearning system that has the capacity to gather and analyze data from learning repositories and to adapt these to the educational curriculum according to the student skills and experience, constructed on advanced machine learning methods [73]. The AEeLS is an inventive work to realistically investigate and recommend relevant educational content based on semantic ontologies techniques. The recommended approach is centered on the successful combination of the OMM and the RMm procedures, which certifies the adaptation of the scheme in the new era learning needs. Also, it suggests a method with a high degree of generalization, by employing a vigorous set of rules qualified to respond to sophisticated education challenges. The implementation of the proposed method was tested on two sophisticated datasets of high complexity. These data sets were selected in order to produce a massive and deep investigation related to the effectiveness of the semantics technologies and specifically with the performance of the ontologies in the educational environment. As proved, the ontologies matching techniques and the recommendations systems are capable to accurately tune in order to solve complicated situations of the modern educational needs. The results have demonstrated the effectiveness of the proposed hybrid method.

*5.2 Innovation*

A momentous novelty of AEeLS is the use of hybrid machine learning methods in order to resolve a multi-dimensional and multi-faceted educational problem. The proposed system mimics in a realistic way the effectiveness of natural knowledge, the practical model of the human brain, and the methods in which the educators' systems use the knowledge, expertise, and experiences.

Also, an essential innovation is the combination of the OMM and the RMm to relocate the expertise of a sophisticated computational decision support system in an eLearning system. This hybrid methodology significantly enriches the way in which the knowledge mining methods work, as it generates the likelihood of forming and combine related content in order to apply knowledge transfer that can be shared with various methods.

Finally, it should not be ignored that a similarly valuable innovation is the fact that the use of AI in order to improve the effectiveness of an educational eLearning system. This improvement expands significantly the way in which the eLearning systems work and respond to the needs of the new education concepts.

*5.3 Future Work*

Forthcoming exploration will concentrate on additional optimization of the parameters that the hybrid system used, in order to achieve faster and more precise results.

Also, further expansion will be achieved by the combination with novel self-improvement and auto-machine-learning methods that can fully automate the identification of relevant educational content.

Additionally, it would be important a comparison study of the performances of the state of the art models in order to investigate the further improvements of our methodology.

Finally, a very vital future enhancement is the upgrading of the method with Natural Language Processing (NLP) capabilities, with Recurrent Neural Network (RNN) and specifically with deep architectures such as Long-Short Term Memory (LSTM), in order to models the time sequences and their dependences with bigger precision and effectiveness.

**Conflicts of Interest**

The authors declare no conflict of interest.

**References**

1. Prinsloo, P., Archer, E., Barnes, G., Chetty, Y., & Van Zyl, D. (2015). Big (ger) data as better data in open distance learning. The International Review of Research in Open and Distributed Learning, 16 (1).
2. Regulation (EU) 2016/679 of the European Parliament and of the Council of 27 April 2016, Available online: https://eur-lex.europa.eu/eli/reg/2016/679/oj, (2020, June 21).
3. Karger, D. R. (2014). The semantic web and end users: What's wrong and how to fix it. Internet Computing, IEEE, 18(6), 64–70.
4. W3C Semantic Web, Available online: https://w3.org/standards/semanticweb/, (2020, June 21).
5. Berners-Lee, T., Hendler, J., & Lassila, O. (2001). The semantic web. Scientific American, 284(5), 34–43.
6. Beydoun, G. (2009). Formal concept analysis for an e-learning semantic web. Expert Systems with Applications, 36(8), 10952–10961.
7. Gerber, A. J., Van der Merwe, A., Barnard, A. (2008). A Functional Semantic Web Architecture. European Semantic Web Conference 2008, ESWC'08, Tenerife, June 2008.
8. Demertzis, K.; Tsiotas, D.; Magafas, L. Modeling and Forecasting the COVID-19 Temporal Spread in Greece: An Exploratory Approach Based on Complex Network Defined Splines. Int. J. Environ. Res. Public Health 2020, 17, 4693.
9. Gooley, A., & Lockwood, F. (Eds.). (2001). Innovation in open and distance learning: Successful development of online and web-based learning. London: Routledge, Taylor & Francis.
10. Masud, M. (2016). Collaborative e-learning systems using semantic data interoperability. Computers in Human Behavior. Vol. 61, pp.127–135.
11. Dabbagh, N., Benson, A. D., Denham, A., Joseph, R., Al-Freih, M., Zgheib, G., Guo, Z. (2016). Massive open online courses. In N. Dabbagh, A. D. Benson, A. Denham, R. Joseph, M. Al-Freih, G. Zgheib, Z. Guo (Eds.), Learning technologies and globalization (pp. 9-13). Heidelberg: Springer International Publishing.
12. Pancerz, K., & Lewicki, A. (2014). Encoding symbolic features in simple decision systems over ontological graphs for PSO and neural network based classifiers. Neurocomputing, 144, 338–345.
13. Kaplan Andreas and Michael Haenlein (2019) Siri, Siri in my Hand, who is the Fairest in the Land? On the Interpretations, Illustrations and Implications of Artificial Intelligence, Business Horizons, 62(1), 15-25
14. Tarus, J.K., Niu, Z. & Mustafa, G. Knowledge-based recommendation: a review of ontology-based recommender systems for e-learning. Artif Intell Rev 50, 21–48 (2018). https://doi.org/10.1007/s10462-017-9539-5
15. Atkinson, J., Gonzalez, A., Munoz, M., & Astudillo, H. (2014). Web metadata extraction and semantic indexing for learning objects extraction. Applied Intelligence, 41(2), 649–664
16. Mao, M., Peng, Y., & Spring, M. (2011). Ontology mapping: As a binary classification problem. Concurrency and Computation: Practice and Experience, 23(9), 1010-1025.
17. Liu, L., Yang, F., Zhang, P., Wu, J.-Y., & Hu, L. (2012). SVM-based ontology matching approach. International Journal of Automation and Computing, 9(3), 306–314.
18. Liu, J., Qin, L., & Wang, H. (2013). An ontology mapping method based on support vector machine. In Proceedings of the 8th International Conference on Ontology Matching-Volume 1111 (pp. 225-226).
19. M. Gaeta, F. Orciuoli, and P. Ritrovato, "Advanced ontology management system for personalised e-Learning," Knowledge-Based Systems, vol. 22, no. 4, pp. 292–301, May 2009, doi: 10.1016/j.knosys.2009.01.006.
20. Yang, K., & Steele, R. (2009). Ontology mapping based on concept classification. 3rd IEEE International Conference on Digital Ecosystems and Technologies, 2009. DEST'09. (pp. 656–661). IEEE.
21. Vidal, J. C., Lama, M., Otero-García, E., & Bugarín, A. (2014). Graph-based semantic annotation for enriching educational content with linked data. Knowledge-Based Systems, 55, 29–42.
22. Luna, J. M., Romero, C., Romero, J. R., & Ventura, S. (2014). An evolutionary algorithm for the discovery of rare class association rules in learning management systems. Applied Intelligence 42(3), 501-513.
23. Peñalver-Martinez, I., Garcia-Sanchez, F., Valencia-Garcia, R., Rodríguez-García, M. Á., Moreno, V., Fraga, A., & Sánchez-Cervantes, J. L. (2014). Feature-based opinion mining through ontologies. Expert Systems with Applications, 41(13), 5995–6008.

24. Wang, J., Peng, J., & Liu, O. (2015). A classification approach for less popular webpages based on latent semantic analysis and rough set model. Expert Systems with Applications, 42(1), 642–648.
25. Yi-Hsing Chang and Hsiu-Yi Huang, "An Automatic Document Classifier System based on Naíve Bayes Classifier and Ontology," 2008 International Conference on Machine Learning and Cybernetics, Kunming, 2008, pp. 3144-3149, doi: 10.1109/ICMLC.2008.4620948.
26. R. Rani and S. Tandon, "Chat Summarization and Sentiment Analysis Techniques in Data Mining," 2018 4th International Conference on Computing Sciences (ICCS), Jalandhar, 2018, pp. 102-106, doi: 10.1109/ICCS.2018.00025.
27. M. A. Khoudja, M. Fareh and H. Bouarfa, "Ontology Matching using Neural Networks: Survey and Analysis," 2018 International Conference on Applied Smart Systems (ICASS), Medea, Algeria, 2018, pp. 1-6, doi: 10.1109/ICASS.2018.8652049.
28. X. -H. Zhi and Y. -F. Li, "Building Ontology Automatically Based on Bayesian Network and PART Neural Network," 2009 WRI Global Congress on Intelligent Systems, Xiamen, 2009, pp. 563-566, doi: 10.1109/GCIS.2009.29.
29. K. Gayathri and A. Marimuthu, "Text document pre-processing with the KNN for classification using the SVM," 2013 7th International Conference on Intelligent Systems and Control (ISCO), Coimbatore, 2013, pp. 453-457, doi: 10.1109/ISCO.2013.6481197.
30. T. Jo, "String Vector based KNN for text categorization," 2018 20th International Conference on Advanced Communication Technology (ICACT), Chuncheon-si Gangwon-do, Korea (South), 2018, pp. 260-265, doi: 10.23919/ICACT.2018.8323718.
31. K. P. P. Shein and T. T. S. Nyunt, "Sentiment Classification Based on Ontology and SVM Classifier," 2010 Second International Conference on Communication Software and Networks, Singapore, 2010, pp. 169-172, doi: 10.1109/ICCSN.2010.35.
32. W. Liu and Q. Li, "Collaborative Filtering Recommender Algorithm Based on Ontology and Singular Value Decomposition," 2019 11th International Conference on Intelligent Human-Machine Systems and Cybernetics (IHMSC), Hangzhou, China, 2019, pp. 134-137, doi: 10.1109/IHMSC.2019.10127.
33. Song Jia, Pu Jiexin and Zhang Ruiling, "The collaborative filtering algorithm based on domain ontology and user preferences," 2012 International Conference on Computer Science and Information Processing (CSIP), Xi'an, Shaanxi, 2012, pp. 902-905, doi: 10.1109/CSIP.2012.6309000.
34. X. Min, Z. Hongfei and Y. Xiaogao, "A Collaborative Filtering Recommendation Algorithm based on Domain Knowledge," 2008 International Symposium on Computational Intelligence and Design, Wuhan, 2008, pp. 220-223, doi: 10.1109/ISCID.2008.139.
35. Aher, S. B., & Lobo, L. M. R. J. (2013). Combination of machine learning algorithms for recommendation of courses in E-Learning System based on historical data. Knowledge-Based Systems, 51, 1–14.
36. Gladun, A., Rogushina, J., García-Sanchez, F., Martínez-Béjar, R., & Fernández-Breis, J. T. (2009). An application of intelligent techniques and semantic web technologies in e-learning environments. Expert Systems with Applications, 36(2), 1922–1931.
37. Wen, C., & Zhang, J. (2015). Design of a microlecture mobile learning system based on smartphone and web platforms. IEEE Transactions on Education, 58(3), 203-207.
38. Katifori, A.; Halatsis, C.; Lepouras, G.; Vassilakis, C.; Giannopoulou, E. (2007). "Ontology Visualization Methods - A Survey" (PDF). ACM Computing Surveys. 39:10. doi:10.1145/1287620.1287621.
39. Li, J., Tang, J., Li, Y., Luo, Q.: Rimom: a dynamic multistrategy ontology align-ment framework. IEEE Trans. Knowl. Data Eng. 21(8), 1218–1232 (2009).
40. Min H., Mobahi H., Vukomanovic S., Irvin K., Krasniqi I., Avramovic S., and Wojtusiak J., "Applying an Ontology-guided Machine Learning Methodology to SEER-MHOS Dataset,", 2016 Bio-ontology at Intelligent Systems for Molecular Biology(ISMB), Orlando, Florida, July 8-9, 2016.
41. Min H., Mobahi H., Vukomanovic S., Irvin K., Krasniqi I., Avramovic S., Wojtusiak J.,"Ontology applications in Machine Learning", 2016 Bio-ontology at Intelligent Systems for Molecular Biology (ISMB), Orlando, Florida, July 8-9, 2016.
42. Tang, J., Li, J., Liang, B., Huang, X., Li, Y., Wang, K.: Using bayesian decision for ontology mapping. Web Semant. 4(4), 243–262 (2006).
43. Euzenat, J., Shvaiko, P.: Ontology Matching, 1st edn. Springer, New York (2007).

15 of 16

44. Mao, M., Peng, Y., Spring, M.: An adaptive ontology mapping approach with neural network based constraint satisfaction. Web Semant. Sci. Serv. Agents World Wide Web 8(1), 14–25 (2010)

45. Isabel, F.P.A., Cruz, F., Stroe, C.: Efficient selection of mappings and auto-matic quality-driven combination of matching methods. In: Workshop on Ontology Matching, pp. 49–60 (2009).

46. Eckert, K., Meilicke, C., Stuckenschmidt, H.: Improving ontology matching using meta-level learning. In: Aroyo, L., Traverso, P., Ciravegna, F., Cimiano, P., Heath, T., Hyv¨onen, E., Mizoguchi, R., Oren, E., Sabou, M., Simperl, E. (eds.) ESWC 2009. LNCS, vol. 5554, pp. 158–172. Springer, Heidelberg (2009).

47. Wang Z. (2014) A Semi-supervised Learning Approach for Ontology Matching. In: Zhao D., Du J., Wang H., Wang P., Ji D., Pan J. (eds) The Semantic Web and Web Science. CSWS 2014. Communications in Computer and Information Science, vol 480. Springer, Berlin, Heidelberg.

48. Zhu, X., Ghahramani, Z., Lafferty, J.: Semi-supervised learning using Gaussian fields and harmonic functions. In:ICML, pp. 912–919 (2003).

49. Bchir O, Frigui H, Ismail MMB (2013) Semi-supervised fuzzy clustering with learnable cluster dependent kernels. International Journal on Artificial Intelligence Tools, 22(3):1-26. Article number 1350013 doi: http://dx.doi.org/10.1142/S0218213013500139

50. Sen P, Namata G, Bilgic M, Getoor L, Galligher B, Eliassi-Rad T (2008) Collective Classification in Network Data. Advancement of Artificial Intelligence, 29(3):93-106.

51. Kecman V. (2001) Learning and Soft Computing. MIR Press. ISBN: 9780262112550

52. Cox E (2005) Fuzzy Modeling and Genetic Algorithms for Data Mining and Exploration. Elsevier Science, USA.

53. Francesco Ricci and Lior Rokach and Bracha Shapira, Introduction to Recommender Systems Handbook, Recommender Systems Handbook, Springer, 2011, pp. 1-35.

54. I. Bartolini, Z. Zhang, and D. Papadias, ``Collaborative filtering with personalized skylines,'' IEEE Trans. Knowl. Data Eng., vol. 23, no. 2, pp. 190203, Feb. 2011.

55. Z. Yang, B. Wu, K. Zheng, X. Wang and L. Lei, "A Survey of Collaborative Filtering-Based Recommender Systems for Mobile Internet Applications," in IEEE Access, vol. 4, pp. 3273-3287, 2016. doi: 10.1109/ACCESS.2016.2573314

56. R. Hu, W. Dou, and J. Liu, ``ClubCF: A clustering-based collaborative fltering approach for big data application,'' IEEE Trans. Emerg. Topics Comput., vol. 2, no. 3, pp. 302:313, Sep. 2014

57. N. Sherif and G. Zhang, "Collaborative filtering using probabilistic matrix factorization and a Bayesian nonparametric model," 2017 IEEE 2nd International Conference on Big Data Analysis (ICBDA) (Beijing, 2017, pp. 391-396, doi: 10.1109/ICBDA.2017.8078847

58. Xiaoyuan Su, Taghi M. Khoshgoftaar, A survey of collaborative filtering techniques, Advances in Artificial Intelligence archive, 2009.

59. Ya-Yueh Shih and Duen-Ren Liu, "Hybrid Recommendation Approaches: Collaborative Filtering via Valuable Content Information," Proceedings of the 38th Annual Hawaii International Conference on System Sciences, 2005, pp. 217b-217b, doi: 10.1109/HICSS.2005.302.

60. Mao, J., Jain, A.K., Duin, P.W.: Statistical pattern recognition: A review. IEEE Trans. on Pattern Analysis and Machine Intelligence 22(1), 4–37 (2000).

61. Fawcett, T., (2006), An introduction to ROC analysis. Pattern Recognition Letters. Elsevier Science Inc. 27(8), 861-874. doi: http://doi.org/10.1016/j.patrec.2005.10.010.

62. Dragisic, Z., Eckert, K., Euzenat, J., Faria, D., Ferrara, A., Granada, R., & Montanelli, S. (2014, October). Results of the ontology alignment evaluation initiative 2014. In P. Shvaiko, M. Mao, J. Li, & A.-C. Ngonga Ngomo (Eds.), Proceedings of the 9th International Conference on Ontology Matching-Volume 1317 (pp. 61-104).

63. ADRIADNE. Available online: http://www.adriadne-eu.org (2020, June 21).

64. MERLOT. Available online: http://www.merlot.org (2016, June 21).

65. Sergio Cerón-Figueroa, Itzamá López-Yáñez, Yenny Villuendas-Rey, Oscar Camacho-Nieto, Mario Aldape-Pérez, and Cornelio Yáñez-Márquez, Instance-Based Ontology Matching For Open and Distance Learning Materials, International Review of Research in Open and Distributed Learning Volume 18, Number 1

66. Anezakis VD., Iliadis L., Demertzis K., Mallinis G. (2017) Hybrid Soft Computing Analytics of Cardiorespiratory Morbidity and Mortality Risk Due to Air Pollution. In: Dokas I., Bellamine-Ben Saoud N., Dugdale J., Díaz P. (eds) Information Systems for Crisis Response and Management in Mediterranean

Countries. ISCRAM-med 2017. Lecture Notes in Business Information Processing, vol 301. Springer, Cham. https://doi.org/10.1007/978-3-319-67633-3_8.
67. Iliadis L., Anezakis VD., Demertzis K., Spartalis S. (2018) Hybrid Soft Computing for Atmospheric Pollution-Climate Change Data Mining. In: Thanh Nguyen N., Kowalczyk R. (eds) Transactions on Computational Collective Intelligence XXX. Lecture Notes in Computer Science, vol 11120. Springer, Cham. https://doi.org/10.1007/978-3-319-99810-7_8.
68. Anezakis VD., Demertzis K., Iliadis L., Spartalis S. (2016) A Hybrid Soft Computing Approach Producing Robust Forest Fire Risk Indices. In: Iliadis L., Maglogiannis I. (eds) Artificial Intelligence Applications and Innovations. AIAI 2016. IFIP Advances in Information and Communication Technology, vol 475. Springer, Cham. https://doi.org/10.1007/978-3-319-44944-9_17
69. Anezakis, V., Demertzis, K., Iliadis, L. et al. Hybrid intelligent modeling of wild fires risk. Evolving Systems 9, 267–283 (2018). https://doi.org/10.1007/s12530-017-9196-6
70. Demertzis Konstantinos, Iliadis Lazaros, Anezakis Vardis-Dimitrios, Commentary: Aedes albopictus and Aedes japonicus—two invasive mosquito species with different temperature niches in Europe, Frontiers in Environmental Science, VOLUME 5, YEAR 2017, PAGES 85, DOI:10.3389/fenvs.2017.00085
71. K. Demertzis, L. Iliadis and V. Anezakis, "A deep spiking machine-hearing system for the case of invasive fish species," 2017 IEEE International Conference on INnovations in Intelligent SysTems and Applications (INISTA), Gdynia, 2017, pp. 23-28, doi: 10.1109/INISTA.2017.8001126.
72. Konstantinos Demertzis, Lazaros S. Iliadis & Vardis-Dimitris Anezakis (2018) Extreme deep learning in biosecurity: the case of machine hearing for marine species identification, Journal of Information and Telecommunication, 2:4, 492-510, DOI: 10.1080/24751839.2018.1501542
73. Demertzis, K.; Kikiras, P.; Tziritas, N.; Sanchez, S.L.; Iliadis, L. The Next Generation Cognitive Security Operations Center: Network Flow Forensics Using Cybersecurity Intelligence. Big Data Cogn. Comput. 2018, 2, 35.